\documentstyle[12pt,epsf]{article}
\def\be{\begin{equation}}
\def\ee{\end{equation}}
\def\bea{\begin{eqnarray}}
\def\eea{\end{eqnarray}}

\def\w{\omega}

\def\gm{\gamma}
\begin{document}
\begin{flushright} 
YITP-97-48 \\ DTP-97-53\\ {\bf hep-th/9710079} 
\end{flushright}
\vskip 1.0 cm

{\Large
\centerline{{\bf Higher-dimensional Generalisations 
of the Euler Top Equations}}
\vskip 1.5cm

\centerline{\qquad \ \ 
  David B. Fairlie{$^\dag$}{$^\ddag$}\footnote{ On research leave 
from the University of Durham, U.K.} and Tatsuya Ueno$^\ddag$\\ }}
\vskip 0.5cm

\centerline{{$^\dag$}Department of Mathematical Sciences,}
\vskip 5pt
\centerline{University of Durham,}
\vskip 5pt
\centerline{Durham, DH1 3LE, England.}
\vskip 5pt
\centerline{{\sl david.fairlie@durham.ac.uk}} 
\vskip 10pt
\centerline{{$^\ddag$}Yukawa Institute for 
Theoretical Physics,}
\vskip 5pt
\centerline{Kyoto University}
\vskip 5pt
\centerline{Kyoto 606-01, Japan.}
\vskip 5pt
\centerline{\quad{\sl tatsuya@yukawa.kyoto-u.ac.jp}} 
\vskip 1.5cm

\begin{center}
{\bf ABSTRACT}
\end{center}
Generalisations of the familiar Euler top equations in three
dimensions are proposed which admit a sufficiently large number  
of conservation laws to permit integrability by quadratures. 
The usual top is a classical analogue of the Nahm equations. 
One of the examples discussed here is a seven-dimensional Euler top, 
which arises as a classical counterpart to the eight-dimensional 
self-dual equations which are currently believed to play a role in 
new developments in string theory.
\newpage

\section{Introduction}
The object of this note is to examine the question of the existence of 
Euler equations in higher dimension which are integrable
generalisations of the familiar Euler equation for  a top, with
quadratic nonlinearities. The motivation for this last requirement 
comes from the fact that the resulting equations may be viewed
as classical versions of the Nahm equations \cite{nahm}.  
The higher-dimensional equivalents of the Nahm equations, i.e. self-dual 
Yang-Mills equations with gauge fields dependent only upon time, which 
have appeared recently in the context of M-Theory 
\cite{cor}\cite{bsk}\cite{hirano}\cite{ckz}\cite{fair}\cite{floratos},
might well be expected to give rise to a similar classical reduction.  
 We shall find that there is indeed an exact parallel, satisfying a
similar Bogomol'nyi bound which is integrable by quadratures, just 
as in the three-dimensional case, but this time it is integrable only in 
principle. There are sufficient conservation laws to permit full
integrability, but their algebraic  solution is prohibitive in practice.
As a by-product a remarkable identity among seven quartic polynomials, 
which reduces the number of conserved quantities to permit a non
constant solution, is discovered. 

\subsection{Three-dimensional example}
 The standard Euler top in three dimensions in its simplest form may
be written as
\be
\dot \omega_1 =\omega_2 \, \omega_3 \ ,
\label{euler3}
\ee
together with cyclic replacements. This set of equations results when
a solution of the three-dimensional Nahm equations,
\be
\frac{dA_i}{dt}=\frac{1}{2}\epsilon_{ijk}[A_j,\ A_k] \ ,
\label{Nahm}
\ee
is sought in the form, in terms of the Pauli matrices $\sigma_i$, 
\be
A_i = \w_i\, \sigma_i\ , \qquad  (i\ {\rm not~~summed.})
\label{notsum}
\ee
{}From (\ref{euler3}), we have two independent conserved quantities,
\be
\w_2^2 - \w_1^2 = c_2 \ , \ \ \ \ \w_3^2 - \w_1^2 = c_3 \ .
                                           \label{eq: cq3}
\ee
As well known, the general solution of (\ref{euler3}) is given 
by elliptic functions. 
Solving (\ref{eq: cq3}) for $\w_2$ and $\w_3$ and 
substituting them into the equation for $\w_1$ in (\ref{euler3}),
we have
\be
\dot{\w}_1 = \sqrt{(\w_1^2 + c_2)(\w_1^2 + c_3)} \ .
\ee
Parametrising the constants $c_2$ and $c_3$ as 
$c_2 = \alpha^2 k^{'2} = \alpha^2 (1 - k^2)$ and $c_3 = \alpha^2$, we have
the solution of the equation,
\be
(\w_1(t), \w_2(t), \w_3(t)) 
= (\alpha k' \, tn(\alpha t + \beta), \alpha k' \, nc(\alpha t + \beta), 
\alpha \, dc(\alpha t + \beta)) \ ,                      \label{eq: gs}
\ee
where $tn(x)$, $nc(x)$ and $dc(x)$ are elliptic functions and 
$\beta$ is a constant.
Note that (\ref{euler3}) is invariant under the change of sign of
any two of $\w_i$ and hence, in addition to the above solution, of
$(+,+,+)$ type say, we have other $(+,-,-)$ and $(-, \pm, \mp)$ type
solutions.

\section{Seven-dimensional top}
A similar procedure applied to the equivalent of the Nahm equations 
for self-dual fields in eight dimensions
\cite{cor}\cite{ckz}\cite{fair}, which is defined using the octonionic
structure constants $C_{ijk}$ $(i,j,k=1,\cdots,7)$ in (\ref{Nahm})
instead of the quaternionic ones $\epsilon_{ijk}$.
For an explicit realisation of the totally anti-symmetric $C_{ijk}$,
\be
C_{127}=C_{631}=C_{541}=C_{532}=C_{246}=C_{734}=C_{567}=1 \ , \qquad
(\rm{ others \ zero})
\ee
we have the set of seven equations;
\bea
&&\frac{d}{dt} A_1- [A_2 ,\ A_7]-[A_6,\ A_3] - [A_5,\ A_4]=0 \ ,
\nonumber\\ 
&&\frac{d}{dt} A_2- [A_7 ,\ A_1]-[A_5,\ A_3] -[A_4,\ A_6]=0 \ ,
\nonumber\\
&&\frac{d}{dt} A_3- [A_1 ,\ A_6]-[A_2,\ A_5] -[A_4,\ A_7]=0 \ ,
\nonumber\\
&&\frac{d}{dt} A_4- [A_1 ,\ A_5]-[A_6,\ A_2] -[A_7,\ A_3]=0 \ ,
\label{orignahm3}\\
&&\frac{d}{dt} A_5- [A_4 ,\ A_1]-[A_3,\ A_2] -[A_6,\ A_7]=0 \ ,
\nonumber\\
&&\frac{d}{dt} A_6- [A_3 ,\ A_1]-[A_2,\ A_4] -[A_7,\ A_5]=0 \ ,
\nonumber\\
&&\frac{d}{dt} A_7- [A_1 ,\ A_2]-[A_3,\ A_4] -[A_5,\ A_6]=0 \ .
\nonumber
\eea
Note that the gauge field in the $t$ direction has been set to zero
to obtain these equations. This generalisation of the Euler top is 
rather different from that proposed by Manakov \cite{manakov}.
A search for a solution in the form
\be
A_i = \w_i\, e_i \ ,  \qquad (i\ {\rm not~~summed,})
\label{notsum2}
\ee
where $e_i$ are a basis for octonions, yields the following equations;
\bea
&&\frac{d}{dt}\w_{1}-\w_{2}\w_{7}-\w_{6}\w_{3}-\w_{5}\w_{4}=0 \ ,
\nonumber\\
&&\frac{d}{dt}\w_{2}-\w_{7}\w_{1}-\w_{5}\w_{3}-\w_{4}\w_{6}=0 \ ,
\nonumber\\
&&\frac{d}{dt}\w_{3}-\w_{1}\w_{6}-\w_{2}\w_{5}-\w_{4}\w_{7}=0 \ ,
\nonumber\\
&&\frac{d}{dt}\w_{4}-\w_{1}\w_{5}-\w_{6}\w_{2}-\w_{7}\w_{3}=0 \ ,
\label{euler7}\\
&&\frac{d}{dt}\w_{5}-\w_{4}\w_{1}-\w_{3}\w_{2}-\w_{6}\w_{7}=0 \ ,
\nonumber\\
&&\frac{d}{dt}\w_{6}-\w_{3}\w_{1}-\w_{2}\w_{4}-\w_{7}\w_{5}=0 \ ,
\nonumber\\
&&\frac{d}{dt}\w_{7}-\w_{1}\w_{2}-\w_{3}\w_{4}-\w_{5}\w_{6}=0 \ .
\nonumber
\eea
The structure of these equations can be read off from Fig.1, the
seven-point plane. This construction arises from the projective
geometry of a plane over a finite field of characteristic two; 3
points lie  on every line and 3 lines pass through each point. 
The contributions to $\dot \w_i$ come from the products of the pairs 
of $\w$'s associated with the other points on each of the three lines 
through $i$.
\begin{figure}[hb]
\epsfxsize= 45 mm
\begin{center}
\leavevmode
\epsfbox{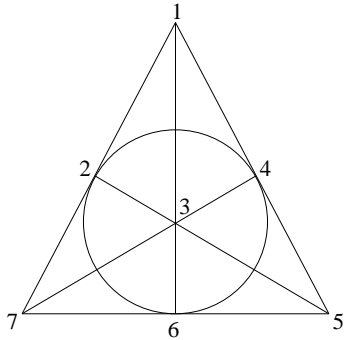}
\end{center}
\caption{7 point plane.}
\end{figure}

\section{Solutions}
This set of equations admits the trivial symmetric solution
\be
\w_i = \frac{-1}{3t} \ ,
\label{trivial}
\ee
and also several broken symmetry solutions, e.g.
\be
\w_1=\w_2=\w_3=\w_4=\frac{-1}{t}\ , \ \ \w_5=\w_6=\frac{1}{t}\ ,
\ \ \w_7=\frac{-3}{t}\ .
\label{broken}
\ee
 The second order equations implied by (\ref{euler7}) come from the 
Lagrangian
\be
{\cal L}=\frac{1}{2}\left(\sum_{i=1}^7(\dot\w_i)^2
+\sum_{i=1}^7\sum_{j<i}\w_i^2\w_j^2\right)
+3\sum_{\rm perms}\w_i\w_j\w_k\w_l \ ,
\label{sondorder}
\ee\
where the last sum in the above expression is taken over all distinct 
values of $\w_i\w_j\w_k\w_l$ corresponding to indices $i,\ j,\ k,\ l$ 
no three of which lie on a line.
The Lagrangian is given up to a divergence by the sum of the squares
of the equations (\ref{euler7}). A similar Bogomol'nyi property is
known for the commutator case \cite{ckz}\cite{matsuo}.

\subsection{Conservation Laws}
The equations (\ref{euler7}) can be rewritten as
\bea
\frac{d}{dt}{(\w_1\pm\w_2)}&=&(\w_6\pm\w_5)(\w_3\pm\w_4)\pm(\w_1\pm\w_2)\w_7 
\ , \nonumber\\
\frac{d}{dt}{(\w_6\pm\w_5)}&=&(\w_3\pm\w_4)(\w_1\pm\w_2)\pm(\w_6\pm\w_5)\w_7
\ ,\label{simpletoo}\\
\frac{d}{dt}{(\w_3\pm\w_4)}&=&(\w_1\pm\w_2)(\w_6\pm\w_5)\pm(\w_3\pm\w_4)\w_7 
\ , \nonumber
\eea
\noindent where either the plus or minus sign is taken consistently
throughout each equation.
The two sets of equations with plus and minus signs are linked by
the seventh equation
\be
\frac{d}{dt}\w_7= \frac{1}{4}\left((\w_1+\w_2)^2+(\w_6+\w_5)^2
+(\w_3+\w_4)^2-(\w_1-\w_2)^2-(\w_6-\w_5)^2-(\w_3-\w_4)^2\right).\label{link}
\ee

Integrals of motion, obtained from these equations are of two types;
\be
\frac{(\w_i\pm\w_j)^2-(\w_k\pm\w_l)^2}{(\w_i\pm\w_j)^2-(\w_m\pm\w_n)^2}
=\alpha_{hikm\pm} \ ,
\label{typeone}
\ee
and
\be
((\w_k+\w_l)^2-(\w_m+\w_n)^2)((\w_k-\w_l)^2-(\w_m-\w_n)^2)=\beta_{hij}
\ , \label{typetwo}
\ee
where $h,\ i,\ j,\ k,\ l,\ m,\ n$ is a permutation of the indices
$1\dots 7$ such that $(i,\ j)$, $(k,\ l)$ and $(m,\ n)$ lie on the
respective three lines  through $h$ in Fig.1 and 
$\alpha_{hikm\pm},\ \beta_{hij}$ are constants of integration.
There are several identities among these relationships, reducing their number.
For example, 
\be
\alpha_{hikm+}\alpha_{hikm-}\alpha_{hmik+}\alpha_{hmik-}\alpha_{hkmi+}
\alpha_{hkmi-}=1 \ , \label{identity}
\ee
and similarly
\be
\alpha_{hikm\pm} = \alpha_{himk\pm}^{-1} \ ,
\qquad  
\alpha_{hikm\pm}+\alpha_{hmki\pm}=1 \ .
\label{simidentity}
\ee
There are still apparently too many. We need to reduce the number of 
independent conserved quantities to six, to enable a proper dynamical 
evolution to take place. This can be achieved as follows, by
concentrating upon the second set, which can be re-expressed as
\be
(\w_k+\w_l+\w_m+\w_n)(\w_k+\w_l-\w_m-\w_n)(\w_k-\w_l+\w_m-\w_n)
(\w_k-\w_l-\w_m+\w_n)=\beta_{hij} \ , 
\label{redundant}
\ee
showing that only 7 of the above  expressions are different. Remarkably, 
the Jacobian of the seven quantities in (\ref{redundant}) with respect
to the variables $\w_j$ is found to vanish, showing that they are
functionally dependent. It is also found, by computer calculations
that the generic rank of this Jacobian is 6, showing that there are 6 
functionally independent conserved quantities. 
The explicit functional relation between the 7 quartic polynomials, 
say $(\gm_1,\gm_2,\gm_3,\gm_4,\gm_5,\gm_6,\gm_7) = 
(\beta_{567},\beta_{631},\beta_{734},\beta_{541},\beta_{532},
\beta_{127},\beta_{246})$,
is given by
\be
\sum_{(i,j,k)}\gm_i\gm_j\gm_k\, \lbrack (\sum_{p=1}^7\gm_p
-2(\gm_i+\gm_j+\gm_k))^2 + \sum_{p=1}^7\gm_p^2 
- 2(\gm_i^2+\gm_j^2+\gm_k^2) \rbrack=0 \ ,
\ee
where the sum is taken over all indices $(i,j,k)$  corresponding to
points lying on lines in Fig.1.
Furthermore, there are no further independent quantities, since any of 
$\alpha_{hikm\pm}$ in the set (\ref{typeone}) can be shown to be a 
solution of one of the following quadratic equations,
\be
\gm_i\, x^2 - (\gm_i + \gm_j - \gm_k) x + \gm_j = 0 \ , 
\ee
where as before, the indices $(i,j,k)$ lie on lines in Fig.1, showing 
that only 6 relations are functionally independent.
Thus we can {\it in principle} solve  for $\w_j,\ j>1$ in terms of 
$\w_1$, substitute in (\ref{euler7}) and thus obtain an implicit
integral solution of this set of equations.  

\section{Other generalisations}
As was noted in the previous section, the equations (\ref{euler7})
reduce to two linked sets of three-dimensional equations. If we set
$\w_7=0$ and remove the final equation, then the set;
\bea
&&\frac{d}{dt}\w_{1}-\w_{6}\w_{3}-\w_{5}\w_{4}=0 \ ,
\qquad
\frac{d}{dt}\w_{2}-\w_{5}\w_{3}-\w_{4}\w_{6}=0 \ ,
\nonumber\\
&&\frac{d}{dt}\w_{3}-\w_{1}\w_{6}-\w_{2}\w_{5}=0 \ ,
\qquad 
\frac{d}{dt}\w_{4}-\w_{1}\w_{5}-\w_{6}\w_{2}=0 \ ,
\label{euler6}\\
&&\frac{d}{dt}\w_{5}-\w_{4}\w_{1}-\w_{3}\w_{2}=0 \ ,
\qquad
\frac{d}{dt}\w_{6}-\w_{3}\w_{1}-\w_{2}\w_{4}=0 \ ,
\nonumber
\eea
decouples into two independent Euler top equations;
\bea
\frac{d}{dt}{(\w_1\pm\w_2)}&=&(\w_6\pm\w_5)(\w_3\pm\w_4) \ ,
\nonumber\\
\frac{d}{dt}{(\w_6\pm\w_5)}&=&(\w_3\pm\w_4)(\w_1\pm\w_2) \ ,
\label{indeuler}\\
\frac{d}{dt}{(\w_3\pm\w_4)}&=&(\w_1\pm\w_2)(\w_6\pm\w_5) \ .
\nonumber
\eea
Thus the system is completely integrable in terms of elliptic integrals.

\subsection{Back to seven dimensions} 
Another seven-dimensional set of equations can be constructed which
provides an interesting example of a system which is partially
integrable, but probably not fully integrable. 

This example is an extension of a simple idea for a generalisation of 
the Euler equations in \cite{dbf}. The idea in that paper is that the 
generalisation to $n$ variables $\omega_i$ is to take the $n$ equations,
\be
\dot \omega_i =\prod^n_{j\neq i}\omega_j \ .
\label{eulern}
\ee
Then it is easy to verify that the quantities,
\be 
\omega_i^2-\omega_j^2=d_j-d_i \ ,
\label{motcons}
\ee
where $d_i$ are constants, are $n-1$ independent constants of  the motion
which can be solved  to express $\omega_j,\ j>1$ in terms of $\omega_1$,
giving a differential equation for $\omega_1$ integrable by
quadratures in terms of hyperelliptic functions, of the form, 
\be
\dot\omega_1=\sqrt{\prod_{j=2}^n(\omega_1^2-d_j)} \ ,
\label{hyper}
\ee
where $d_1$ has been set to zero for aesthetic reasons.
Now this result can be used to find an integrable subset of solutions
of a related set of equations with quadratic nonlinearities, namely
\be
\frac{d}{dt}y_{ij}=\sum^n_{k\neq i,j}y_{ik}y_{jk} \ .
\qquad \ \ (i\not=j)
\label{relate} 
\ee
Through the ansatz 
\be
y_{ij}= \prod^n_{k\neq i,j}\omega_k \ , \qquad \qquad \ \ (i\not=j)
\label{ansatz}
\ee
it may be readily verified that a solution to (\ref{relate}) is given 
in terms of the solution to (\ref{eulern}), which has been already 
discussed in \cite{dbf}.
The $n=4$ case in (\ref{relate}), with $y_{ij}=y_{ji}$, is
exceptional. This is nothing but the set of equations (\ref{euler6})
and is completely integrable. Hence general solutions of four-dimensional 
Euler top equations in (\ref{eulern}) are given by elliptic
functions.
\par       

The number of variables for which this stratagem works is 
$\frac{1}{2}n(n-1)$. What we want to do is to realise a similar
connection for seven-dimensional equations. 
Consider the equations of motion;
\bea
\frac{d}{dt}y_{1}&=&y_{2}y_{4}+y_{6}^2+y_{3}y_{7} \ ,
\nonumber\\
\frac{d}{dt}y_{2}&=&y_{5}y_{1}+y_{3}^2+y_{6}y_{7} \ ,
\nonumber\\
\frac{d}{dt}y_{3}&=&y_{4}y_{6}+y_{2}^2+y_{5}y_{7} \ ,
\label{euler9}\\
\frac{d}{dt}y_{4}&=&y_{1}y_{3}+y_{5}^2+y_{2}y_{7} \ ,
\nonumber\\
\frac{d}{dt}y_{5}&=&y_{6}y_{2}+y_{4}^2+y_{1}y_{7} \ ,
\nonumber\\
\frac{d}{dt}y_{6}&=&y_{3}y_{5}+y_{1}^2+y_{4}y_{7} \ ,
\nonumber\\
\frac{d}{dt}y_{7}&=&y_{1}y_{2}+y_{3}y_{4}+y_{5}y_{6} \ .
\nonumber
\eea
Then the following slight extension of the three-dimensional Euler 
equations;
\be
\dot \omega_1 =\omega_2^2 \omega_3^2 \ ,
\label{euler3a}
\ee
together with cyclic replacements can be related to the equations 
(\ref{euler9}) by defining the following seven quantities,
\be
y_{1}=\w_1^2\w_2\ , \ y_{3}=\w_2^2\w_3\ , \ y_{5}=\w_1\w_3^2\ , \ 
y_{2}=\w_1\w_2^2\ , \ y_{6}=\w_1^2\w_3\ , \ y_{4}=\w_2\w_3^2\ , \ 
y_{7}=\w_1\w_2\w_3 \ . \label{seven} 
\ee
A solution of (\ref{euler3a}) will give rise to a solution of the set
of seven equations with quadratic nonlinearities (\ref{euler9}). Such
a solution may  be constructed by quadratures since it is
easy to see that
\be
 \w_2^3 =\w_1^3-c_2\ , \ \ \w_3^3 =\w_1^3-c_3 \ , \ \ 
(c_2,\ c_3,\ {\rm constant})
\label{newcons}
\ee
and hence the equations can be solved in terms of $\w_1$ and thus, 
in principle the equations (\ref{euler3a}) are reduced to integrations.
We have been able to find only one quartic conserved quantity for the set
(\ref{euler9}), namely
\bea
&&(- y_1 y_4 + y_2 y_5 + y_3 y_6 - y_7^2)^2 
- 4 (y_3 y_5 - y_4 y_7) (y_2 y_6 - y_1 y_7) = 
\nonumber \\
&&\ \ (y_1 y_4 + y_2 y_5 - y_3 y_6 - y_7^2)^2 
- 4 (y_2 y_4 - y_3 y_7)(y_1 y_5 - y_6 y_7)  = 
\nonumber \\
&&\ \ (y_1 y_4 - y_2 y_5 + y_3 y_6 - y_7^2)^2
- 4 (y_4 y_6 - y_5 y_7)(y_1 y_3 - y_2 y_7) = d \ ,
\eea
and it appears likely that the full system is not integrable, when the 
constraints implied by (\ref{seven}), namely
\be
y_1y_4\ =\ y_2y_5\ =\ y_7^2\ ,  \qquad y_1y_3\ =\ y_2y_7\ ,
\qquad y_2y_6\ = y_1y_7 \ ,
\label{constraints}
\ee
are relaxed.

\section*{Acknowledgements}
D. Fairlie thanks the Yukawa Institute for Theoretical Physics,
Kyoto for their hospitality and the financial support of both 
the Royal Society of London and the Japan Society for the Promotion of 
Science (JSPS). 
T.Ueno is supported by the JSPS, No.\,6293.

\end{document}